\documentclass[11pt,a4paper]{article}
\pdfoutput=1
\usepackage{jheppub}
\usepackage[T1]{fontenc}
\usepackage{epsfig}
\newcommand{\bu}{$\bullet$\ }
\newcommand\blue{\color{blue}}

\newcommand{\be}{\begin{equation}}
\newcommand{\ee}{\end{equation}}
\newcommand{\bes}{\begin{subequations}}
\newcommand{\ees}{\end{subequations}}
\newcommand{\bea}{\begin{eqnarray}}
\newcommand{\eea}{\end{eqnarray}}
\newcommand{\bear}{\begin{equation}\begin{array}}
\newcommand{\eear}[1]{\end{array}\label{#1}\end{equation}}
\usepackage{wrapfig}

\def\ba{$$\begin{array}}
 \def\ea{\end{array}$$}

\newcommand{\fr}[2]{\dfrac{{ #1}}{{ #2}}}

\newcommand{\fn}[1]{\footnote{{#1}}}

\def\vep{{\varepsilon}}

\newcommand{\ggam}{\mbox{$\gamma\gamma\,$}}

{\end{list}}
\newcounter{enumct}

\begin{document}

\title{Two-photon Higgs width and triple Higgs coupling in 2HDM at SM-like scenario}
\author[a,b]{I.~F.~Ginzburg}
\affiliation[a]{ Sobolev Institute of Mathematics\\ Novosibirsk, 630090, prosp.ac.Koptyug,4, Russia} \affiliation[b]{Novosibirsk State University\\ Novosibirsk, 630090, Pirogova str.,2, Russia}
\emailAdd{ Ginzburg@math.nsc.ru}

\abstract{
Within 2HDM, sizable deviations of the  triple Higgs coupling and the two-photon Higgs width from their SM values can
have a common origin. If SM-like scenario for the observed Higgs boson is realized, mentioned deviations can be either visible simultaneously or  not observable at the LHC.}

\maketitle
\flushbottom


\section{Introduction}

The recent discovery of a Higgs boson with $M\approx 125$~GeV at the LHC \cite{125Higgs-1}-\cite{125Higgs-4} suggests that the spontaneous electroweak symmetry breaking is most probably brought up by the Higgs mechanism. The simplest realization of the Higgs mechanism introduces a single scalar isodoublet $\phi$ with the Higgs potential $V_H=-m^2(\phi^\dagger \phi)/2 +\lambda(\phi^\dagger \phi)^2/2$. This model is usually called ``the Standard Model'' (SM).

The current data do not rule out the possibility of extended Higgs models. The two-Higgs-Doublet model (2HDM) presents the simplest extension of the standard Higgs model \cite{TDLee}. Here the standard Higgs doublet is supplemented by an extra hypercharge-one doublet. The 2HDM offers a number of phenomenological scenarios with different physical content in different regions of the model parameter space, such as a natural mechanism for spontaneous CP violation, etc. \cite{TDLee}-\cite{Branco2HDM}. For example, the  Higgs sector of the MSSM is a particular case of 2HDM. Some variants of 2HDM have interesting cosmological consequences \cite{GIK09-1},\cite{GIK09-2}.

This model contains a charged Higgs boson $H^\pm$ with mass $M_\pm$ and three neutral Higgs scalars $h_a$ with masses $M_a$, generally with indefinite CP parity. We name the observed Higgs boson as $h_1$ ($M_1=125$~GeV). The observation of all these particles is a necessary step in the verification of the model. However, new bosons $h_{2,3}$, $H^\pm$ may escape observation in the experiments of nearest future (see for details \cite{Gin15rep-1},\cite{Gin15rep-2} and many references therein).
Nevertheless, even before this observation, there are processes  that allow in principle to detect some  effects beyond the SM.

{\it (i)} The detection of a triple  Higgs vertex $g(h_1h_1h_1)$, which is scheduled in the LHC and other colliders, presents the necessary step in the verification of the Higgs mechanism.
 The main problem discussed below is -- in which cases future observation of this vertex will be useful
for finding  New Physics effects. We note that it will be useful  only if  this vertex differs strongly from its SM value. We discuss cases in which it can happen provided SM-like scenario for Higgs boson (see below) is realized. One important opportunity is related to the  unusual value of $H^+H^-h$ interaction. If such value is realized, visible effect in the two-photon width should be observed. That is why we include in the text the discussion of this width.

{\it (ii)} The  loop obliged  interactions of the Higgs boson ($h_1\ggam$, $h_1Z\gamma$, $h_1gg$) present a good opportunity to discover unusual properties of $h_1$ in its interactions with other particles. The value of $h_1\ggam$ interaction depends  on the value of the $H^+H^-h_1$ vertex regardless of the possibility {\blue of observing} $H^\pm$.
These loop obliged couplings were calculated in many papers. However, the effects of a charged Higgs boson loop and the possible admixture of CP-odd component of the neutral Higgs boson  have not been analyzed in sufficient detail until now.

\section{Basics}

{\bf Relative couplings.}

We   use   relative couplings, defined as ratios of the couplings of each neutral Higgs boson $h_a$   with the fundamental particle $P$ to the corresponding SM couplings and dimensionless relative couplings for  interactions $H^\pm W^\mp h_a$  and $H^+H^-h_a$:
\bear{c}
\chi^P_{a}=\fr{g^P_a}{g^P_{\rm SM}} \qquad (P=V\,(W,Z) , q=(t,b,...), \ell=(\tau,...))\,;\\[2mm]
    \chi_{a}^{H^+ W^-} = \dfrac{g(H^+ W^- h_a)}{M_W/v}\,,\qquad
\chi^\pm_a=\fr{g(H^+H^-h_a)}{2M_\pm^2/v}\,.
\eear{relcoupl}

The neutrals $h_a$  generally have no definite CP parity. Couplings $\chi^V_a$ and $\chi^{\pm }_a$ are real due to Hermiticity of Lagrangian, while other couplings  are generally complex.
The $Re(\chi_a^f)$ and $Im(\chi_a^f)$ are responsible for the interaction of fermion $f$ with CP-even and CP-odd parts of $h_a$ respectively. (In particular, for the CP-conserving case with $h_3=A$ we have $Im(\chi_{2,1}^f)=0$, $Re(\chi_3^f)=0$).

The relative couplings obey the following {\bf sum rules} \cite{gunion-haber-wudka}-\cite{GKan}:
 \be
\sum\limits_a (\chi^V_{a})^2=1\,,\qquad
|\chi_{a}^V|^2+| \chi_{a}^{H^\pm W^\mp}|^2=1
\,,\qquad \sum\limits_a (\chi^f_a)^2=1\,.\label{SRV}
 \ee

We  omit the adjective "relative" further in the text.\\

{\bf The minimal complete set of measurable quantities.}

The minimal complete set of measurable quantities (we call them "observables") determines  all parameters of the most general Higgs Lagrangian. It was found in \cite{GKan}. This set is subdivided naturally into two subsets.  The first subset contains v.e.v. of Higgs field $v=246$~GeV, masses of all Higgs bosons $M_{1,2,3}$, $M_\pm$ and two out of three couplings $\chi_a^V$. To form the second subset of complete set, one  need to use triple and quartic Higgs self-interactions.
The minimal simple collection of parameters of the second subset contains three triple couplings $H^+H^-h_a$ (quantities $\chi^\pm_a$) and one quartic coupling $g(H^+H^-H^+H^-)$.
The  parameters of Higgs potential are expressed simply via these observables   \cite{GKan}.

In the most general 2HDM, all these observables are independent of each other. Their possible values are only limited by general conditions, such as positivity and sum rules \eqref{SRV}. In some special variants of 2HDM,  additional relations between these parameters may appear (for example, in the CP conserving case  we have  $\chi^V_3=0$, $\chi_3^\pm=0$).\\

{\bf SM-like scenario.}

Experimental data allow to suggest
 that the Nature realizes SM-like scenario:

1)  We observe a single Higgs boson. Its mass $M\approx 125$~GeV, we call it  $h_1$.

2) The Higgs boson couplings with fundamental particles $P$ (gauge bosons $V$ and fermions $f$) are close to the SM  expectations within experimental accuracy (see e.g. \cite{125_2HDM-1}-\cite{125_2HDM-3}):
 \be
\vep_P=\left|1-|\chi^P_1|^2\right|\ll 1\,\qquad (P=V(W,\,Z),\;\;\; f=(t,\,b,\,\tau,...))\,.\label{SMlike}
 \ee
However, this statement remains only a plausible hypothesis
until  the couplings are measured with sufficient accuracy.
The crucial point of forthcoming discussion is a small, but non-zero, value of $\vep_V$, for example $\vep_V\sim 0.1$. (The {\it  decoupling limit} is realized at $\vep_V\to 0$.)
The $W$-fusion experiments are of the greatest interest here.\\

{\bf Some couplings in the SM-like scenario}

1. Because  of the first SR \eqref{SRV}, the couplings of  other neutrals $h_a$ with gauge bosons $\chi^V_a$ are small (these Higgses are {\it gaugefobic}),
  \be
  |\chi^V_a|^2<\vep_V\ll 1\,, \quad a=2,\,3\,.\label{chVa}
  \ee

2. Because of the second  SR \eqref{SRV}, the absolute values of non-diagonal couplings  with EW gauge bosons $\chi^{W^\pm H^\mp}_a$ for $a=2,\,3$ are close to their maximal values, while similar coupling for the observed Higgs $\chi^{W^\pm H^\mp}_1$ is small\fn{The calculations of $H^-\to W^-h_1$ decay at LHC in \cite{WHh1-1},\cite{WHh1-2} are made in the particular case of CP-conserving 2HDM  and in addition to that, in the case when $\vep_V$ is not very small.
}:
\be
a) \;\;|\chi^{W^\pm H^\mp}_a|^2\approx 1\,;\quad b)\;\;
|\chi^{W^\pm H^\mp}_1|^2\sim \vep_V\ll 1\,.\label{chWa}
\ee

\section{Higgs two-photon width and gluon fusion} \label{secggam}

The equations for Higgs two photon width $\Gamma(h_a\to \ggam)$
were originally derived in \cite{Vain}. These widths are   the sums of C-even and C-odd contributions:
  \be
\Gamma^{\gamma\gamma}_a =\fr{\alpha^2M_a^3}
{256\pi^3 v^2}\left(|\Phi^{E\gamma}_a|^2+|\Phi^{O\gamma}_a|^2\right)\,.
\ee
In turn,  quantities $\Phi^{E\gamma}_a$ and $\Phi^{O\gamma}_a$ are the sums of the contributions $\Phi_J(r^P_a)$ of different charged particles $P$ with mass $M_P$ and spin $J$, circulating in loops (superscript E and O mark CP-even and CP-odd quark loop contributions).
\bear{c}
\Phi^{E\gamma}_a=\chi^V_a\Phi_1(r^W_a)+\sum\limits_f Re\chi^f_a
N_c Q_f^2\; \Phi^E_{1/2}(r^f_a)+
\chi^\pm_a\Phi_{0}(r^\pm_a)\,,\\[2mm]
\Phi^{O\gamma}_a=\sum\limits_f Im\chi^f_a
N_c Q_f^2
\Phi^O_{1/2}(r^f_a)\,;\qquad r^P_a=\fr{4M_P^2}{M_a^2}\,.
\eear{ggamwid}
In contrast to the SM, in these equations the contribution of $H^\pm$ (not discovered yet) is added. (At large    $|\chi_a^t|$ and (or)  $|\chi_a^\pm|$  one-loop equations $\Phi_{1/2}^{E,O}$ and $\Phi_0$ must be modified.)

\bear{c}
\Phi_{1} (r) = 2+3r+3r(2-r)\phi^2(r)\,,\quad
\Phi_{0}(r) = r[1-r\phi^2(r)]\,,\\[2mm]
 \Phi^E_{1/2}(r) =
-2r[1+(1-r)\phi^2(r)]\,,\quad
\Phi^O_{1/2}(r) = -2r\phi^2(r)\,.
 \eear{ggamfunc}

\be
 \phi(r)=\theta(r-1)\arcsin\fr{1}{\sqrt{r}}+\theta(1-r)
\left(\fr{\pi}{2}\,\theta(r)+i\ln\left(\fr{1+\sqrt{1-r}}{\sqrt{|r|}}\right)
\right)\,.\label{phidef}
\ee

The latest data show that $\Gamma(h_1\to \ggam)$ is close to its SM value. In assuming   $h_1$ to be CP-even and  $|\chi_1^\pm|\lesssim 1 $,  these observations provide a basis for the claim that
$\chi_1^V\approx 1$  and $\chi_1^t\approx 1$ (SM-like scenario) \cite{125_2HDM-1}- \cite{125_2HDM-3}. This very opportunity was considered  in \cite{GKO-1}-\cite{GKO-5} where two  facts about Higgs with mass $M_1= 100\div 140$~GeV were established.\\ {\it(i)} Contribution from charged Higgs loop  with $\chi_1^\pm \approx 1$ reduces $\Gamma(h_1\to \ggam)$  by about 10\% (that is within accuracy of modern data).\\ {\it(ii)} At $\chi_1^t\approx-1$ the width $\Gamma(h_1\to \ggam)$ increases by factor about 2.5.

The latter fact means that the value of $\Gamma(h_a\to \ggam)$, close to SM value, can be obtained not only at $\chi_1^t\approx1$ but also at negative $\chi_1^t$ with $|\chi_1^t|<1$. Similar conclusions were obtained in detailed analysis of modern data  in \cite{125_2HDM-1}-\cite{125_2HDM-3}.
The cases of  big CP-odd admixture in $h_1$  and sizable difference $\chi_1^\pm-1$ are not explored in details yet. In particular, they were left out in the interpretation of data \cite{PexpCMS1}-\cite{PexpATLAS2} -- see \cite{Higgspar}.\\

{\bf Gluon fusion.}

The cross section of gluon fusion is given by the same quark loop integrals \eqref{ggamfunc}. This cross section is saturated by  $t$-quark loop. Therefore in the one-loop approximation this cross section is expressed via  the cross section $\sigma(gg\to h^{wb}_{SM}(M_a))$ for possible SM Higgs boson with mass $M_a$:
\bear{c}
\sigma(gg\to h_a)=\sigma(gg\to h^{wb}_{SM}(M_a))\left[(Re\chi^t_a)^2
+(Im\chi^t_a)^2\Phi^{O/E}(r^t_a)\right],
\\[2mm] \;\;where \;\;\;\;
\Phi^{O/E}(r)=
\left(\Phi^O_{1/2}(r)/\Phi^E_{1/2}(r)\right)^2\,.
\eear{glufus}
For $M_a=125$~GeV and 300~GeV we have $\Phi^{O/E}\approx 2.25$ and $2.7$ respectively.

\section{Triple Higgs coupling}\label{sechhh}

The measuring of $g(h_1h_1h_1)$ is scheduled in the LHC and other colliders. The accuracy of these measurements can not be high,  since in each case corresponding experiments deal with interference of two channels with identical final state -- an independent production of two Higgses and production of Higgses via $h_1h_1h_1$ vertex (at LHC  -- from $t$-loop). This interference is mainly destructive \cite{hhhinterf}. For example, for 100~TeV hadron collider with total luminosity 3/{\it abn} one can hope to reach accuracy  of 40\% in the extraction of this vertex from future data \cite{hhhestim}.

The equation for triple Higgs coupling in terms of the introduced observables in the most general 2HDM was found in the \cite{GKan}:
\bear{c}
g(h_1h_1h_1)=\fr{M_1^2}{v}\chi_{111};\quad
\chi_{111}=
\chi^V_1\left\{1+\!\left(1-(\chi^V_1)^2\right)\left[1
+\sum\limits_b 2\fr{M_b^2}{M_1^2}(\chi^V_b)^2\right]\right.+\\[2mm]
\left.+\!\left(1-(\chi^V_1)^2\right)\fr{2M_\pm^2}{M_1^2}\left[\sum\limits_b\chi^V_b\chi^\pm_b-1+Re\left(\sum\limits_b
\chi^{H^+W^-}_b\chi^\pm_b\fr{\chi^{H^+W^-}_1}{\chi^V_1}\right)\right]\right\}\,.
\eear{triple1-c}
Here factor $M_1^2/v$ is the SM result, and $\chi_{111} -1$ represents the New Physics effect.

In the  SM-like scenario  it is easy to estimate
\bear{c}
\chi_{111}\approx
(1-\vep_V/2)\left\{1+\vep_V\left[3+B\vep_V +2B_\pm\left(\chi_1^\pm -1 +\vep_V K_\pm\right)\right]\right\}\,,\\[2mm]
B\sim \sum\limits_b M_b^2/M_1^2; \quad B_\pm=2M_\pm^2/M_1^2\,,\quad K_\pm \sim \chi_b^\pm\,,(b=2,\,3)\,.
\eear{hhhappr}
We see that at  moderate values of parameters, relative coupling $\chi_{111}$ is close to 1, and it is difficult to expect sizable effect\fn{For the particular CP conserving case and with moderate values of parameters such conclusion was obtained in \cite{hhhviol}, \cite{hhhviol1} (see also \cite{hhhMSSMn-1}, \cite{hhhMSSMn-2} for  the CP conserving MSSM).  For the nMSSM (2HDM +Higgs singlet) values $\chi_{111}$ can vary from  1.9 to -1.1 \cite{hhhMSSMn-1}, \cite{hhhMSSMn-2}.},\fn{For some particular variant of MSSM the value of triple Higgs coupling with radiative correction $g^{ren}(h_1h_1h_1)$ looks essentially different from its tree form in SM, $M_1^2/v$ \cite{hhhRC}. However, in this very approximation one must take into account the mass renormalization  $M_1\to M_1^{ren}$. In ref.~ \cite{Boudhhh} it was found  that
 $g^{ren}(h_1h_1h_1)\approx (M_1^{ren})^2/v$ -- similar to  the SM. }.

There are {\it special exotic  values of model parameters} providing sizable deviations of triple Higgs coupling from its SM value, i.e. $|\chi(h_1h_1h_1)-1|\gtrsim 1$.  Our estimates are valid if $\vep_V$ is not extremely small. (In the opposite case  big deviations of $\chi_{111}$ from 1 can appear in the range of parameters, violating perturbativity  and giving partially strong interaction in the Higgs sector with possible new phenomena. This case  should be explored separately.)
For numerical estimations we use   $\vep^V_1\approx 0.1$.

{\it (i)} The most interesting case presents itself if the value of $B_\pm\;\chi_1^\pm$ product is big ($\gtrsim 1/\vep_V$). The big value of vertex $H^+H^-h_1$ (even at moderate value of charged Higgs mass) results in $|\chi(h_1h_1h_1)-1|\gtrsim 1$. Simultaneously it gives big effect in  $\Gamma(h_1\to\ggam)$. Coexistence of these two phenomena  can be an important source of knowledge about the charged Higgs boson before its direct discovery.

{\it (ii)} Other non-trivial opportunities for observation of sizable effect in triple Higgs coupling present themselves in
less natural cases -- when
 quantities $B$ and (or) $B_\pm K_\pm$ are huge, $\gtrsim 1/\vep_V^2$:

{\it (ii-a)} One or both of Higgs neutrals $h_{2,3}$ are  heavier than a few TeV. Direct discovery of such Higgs seems to be a difficult task. Therefore  for a long time detection of this phenomenon may become an important source of knowledge about these heavy neutrals.

{\it (ii-b)} The  couplings $\chi_a^\pm\gtrsim 10$.  In this case the  two-photon width $h_a\to\ggam$ will be strongly different from similar width, calculated for the would-be SM Higgs boson with the same mass.

\section{Summary}\label{secsum}

The presented discussion can be summarized in the following points.

\bu The non-trivial effects can appear only in the case when $\vep_V$ is not extremely small.

\bu {\bf Two photon width $\pmb{\Gamma(h_1\to\ggam)}$}. Direct measuring of $h_1WW$ and $h_1t\bar{t}$ couplings is a very important task. Besides, the recent analysis of two photon width and gluon fusion for observed Higgs boson $h_1$ should be supplemented by more detailed study of effect of possible CP violation. In this respect experimental study of asymmetries in the $h_1\to\tau\bar{\tau}$ etc. is essential (see  \cite{CPtau} for estimates). Unfortunately,
high accuracy in the determination of $h_1H^+H^-$ coupling will not be achieved in the nearest future, because of low accuracy in the measuring $h_1\ggam$ vertex at LHC.

\bu {\bf Triple Higgs vertex}. At moderate values of masses, etc. the observation of sizable deviation from SM prediction in the triple Higgs vertex is unlikely. The sizable effect here may occur in more or less exotic cases.\\ {\it(i)}  Strong enough $h_1H^+H^-$ interaction (it gives an effect  in $\Gamma(h_1\to \ggam)$).\\
 {\it(ii-a)}  Extremely heavy additional neutral Higgs bosons.\\
 {\it(ii-b)} Strong interaction $h_aH^+H^-$.

\bu \ Special case appears in the SM-like scenario at 400~GeV$>M_2>250$~GeV if $|\chi^t_2|>1$. In this case Higgs boson $h_2$ is relatively narrow and the cross section of gluon fusion $gg\to h_2$ can be  larger than that for the would-be SM Higgs boson with mass $M_2$. The process $gg\to h_2\to h_1h_1$ can be seen as a resonant production of $h_1h_1$ pair. In principle, it allows to discover the mentioned $h_2$ at LHC (see example in \cite{hhhviol1}, \cite{H+H-h_1}, \cite{h2hh} for special sets of parameters).

\acknowledgments

The discussions with F. Boudjema, I.Ivanov, K. Kanishev, M. Krawczyk, P.Osland, M.~Vysotsky were useful.
This work was supported in part by grants RFBR  15-02-05868, NSh-3802.2012.2 and  NCN OPUS 2012/05/B/ST2/03306 (2012-2016).\\

{Higgspar}

\end{document}